%% file: arxiv.tex
\documentclass[10pt]{article}

\usepackage[utf8]{inputenc}
\usepackage{lmodern}
\usepackage{amsmath}
\usepackage{amssymb}
\usepackage{color}
\usepackage{array}
\usepackage{float}
\usepackage{graphicx}
\usepackage{algorithm2e} \RestyleAlgo{ruled}
\usepackage[super]{nth}
\usepackage[labelfont=bf]{caption}
\usepackage{hhline}
\usepackage{cleveref}
\usepackage{tipa}
\crefname{figure}{Figure}{Figures}
\Crefname{figure}{Figure}{Figures}
\crefname{table}{Table}{Tables}
\Crefname{table}{Table}{Tables}
\crefname{section}{Section}{Sections}
\Crefname{section}{Section}{Sections}
\crefname{algorithm}{Algorithm}{Algorithms}
\Crefname{algorithm}{Algorithm}{Algorithms}
\usepackage{listings}
\lstdefinestyle{mystyle}{
    basicstyle=\ttfamily\footnotesize,
    breakatwhitespace=false,
    breaklines=true,
    keepspaces=true,
    showspaces=false,
    showstringspaces=false,
    showtabs=false}
\usepackage[backend=bibtex,style=phys,biblabel=brackets,natbib,sorting=nty]{biblatex}
\title{ 
GEMMFIP: Unifying GEMM in BLIS}

\author{RuQing G. Xu\\
\normalsize\it Department of Physics \\
\normalsize\it
The University of Tokyo \\
\normalsize\it
Hongo 7-3-1, Bunkyo-Ward, Tokyo 113-0033, Japan \\
\normalsize
{\tt ruqing.xu@phys.s.u-tokyo.ac.jp} \\[0.2in]
Field G. Van Zee \\
Robert A. van de Geijn \\
\normalsize\it
Department of Computer Science \\
\normalsize\it
\& \\
\normalsize\it
Oden Institute \\
\normalsize\it
The University of Texas at Austin \\
\normalsize\it
Austin, TX 78712, U.~S.~A. \\
\normalsize
{\tt field,rvdg@cs.utexas.edu}
}

\date{\today}
\newcommand{\gemm}{\textsc{gemm}}
\newcommand{\gemmsup}{\textsc{gemmsup}}
\newcommand{\gemmfip}{\textsc{gemmfip}}

\newcommand{\arma}{\textsc{Arm}}
\newcommand{\arm}{\textsc{Arm} NEON}
\newcommand{\armsve}{\textsc{Arm} SVE}
\newcommand{\ia}{x86\_64}
\newcommand{\lda}{\text{LDim}}
\newcommand{\bigo}{\mathcal{O}}

\begin{document}

\maketitle

\begin{abstract}
Matrix libraries often focus on achieving high performance for problems considered to be either ``small'' or ``large'', as these two scenarios tend to respond best to different optimization strategies.
We propose a unified technique for implementing matrix operations like general matrix multiplication (\gemm) that can achieve high performance for both small and large problem sizes.  
The key is to fuse packing -- an operation that copies data to a contiguous layout in memory and which is critical for large matrix performance -- with the first computational ``pass'' over that data.
This boosts performance across the problem size spectrum.
As a result, tuning general-purpose libraries becomes simpler since it obviates the need to carefully express and parameterize logic that chooses between a ``small matrix'' strategy and a ``large matrix'' strategy.
A prototype implementation of the technique built with the BLAS-like Library Instantiation Software (BLIS) framework is described and performance on a range of architectures is reported.
\end{abstract}

\section{Introduction}
The Basic Linear Algebra Subprograms (BLAS)~\cite{BLAS1,BLAS2,BLAS3} interface has had a profound impact on scientific software development. 
It is now also of great importance to fields like machine learning and data analytics.
By coding applications in terms of the BLAS, portable high performance can be achieved.  For this reason, whenever a new high-performance computer architecture arrives, the instantiation of this interface is a high priority.

Historically, it was expected that vendors leverage their expertise with the architecture to create proprietary matrix libraries, with key components coded in assembly language.
IBM's algorithms and architectures approach demonstrated that by co-designing architectures, compilers, and libraries it was possible to achieve high performance with implementations coded in a high-level language (Fortran)~\cite{IBM:P2}.
This inspired a number of open-source efforts to provide portable implementations of the BLAS, including
the Automatic Tuned Linear Algebra Software (ATLAS)~\cite{ATLAS2}, 
the GotoBLAS~\cite{Goto1,Goto2}, 
the OpenBLAS~\cite{OpenBLAS} (a fork of GotoBLAS), and the BLAS-like Library Instantiation Software (BLIS)~\cite{BLIS1,BLIS2} upon which this paper implements its approach.
An added benefit of BLIS is that it supports an analytical model for determining blocking parameters so that autotuning can be avoided~\cite{10.1145/2925987}.

Across all publically available efforts towards matrix libraries, a fundamental problem that complicates the implementation of matrix-matrix operations, a.~k.~a.~level-3 BLAS, is that packing to improve data locality, which is necessary for high performance when targeting large matrix sizes, actually \emph{impedes} high performance for smaller matrix sizes. 
It has been thought that this is an inherent problem that can only be solved by implementing separate code paths for small and large matrix sizes and then selecting one of them based on the problem size characteristics.
In this paper, we provide preliminary evidence that this conventional wisdom \emph{may be wrong}:  the two code paths can be unified in a way that mostly preserves the benefits of both.  This is achieved by integrating packing -- which is optional in the small code path -- more tightly with the computation.  Importantly, the problem sizes where optional packing should be turned on or off can be more easily justified and encoded.

\section{Goto's Algorithm and its Instantiation in BLIS}
\label{sec:BLIS}

Goto's algorithm~\cite{GotoTR,Goto1,Goto2} was first developed for CPUs with two levels of cache and continues to be the algorithm that underlies most if not all vendor and open-source implementations of the level-3 BLAS.  This section gives a high-level description of this algorithm and its instantiation in BLIS.

\subsection{Goto's algorithm for large matrices}
\begin{figure}[htb!]
\centering
\includegraphics[width=0.75\textwidth]{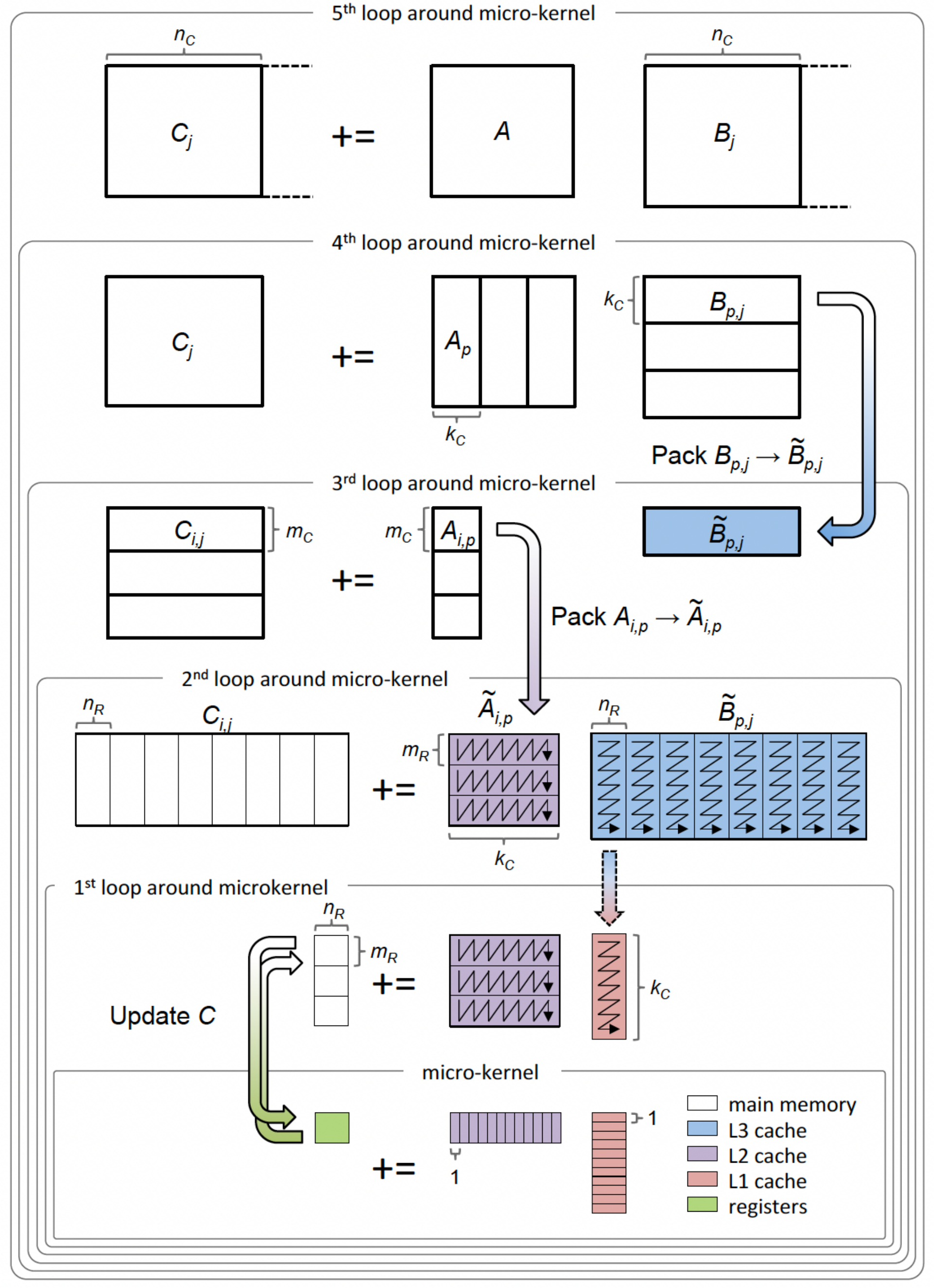}
\caption{
Goto's algorithm for \gemm\ as five loops around the microkernel.
This diagram, which is often used when explaining the fundamental techniques that underly the BLIS implementation of \gemm, was modified from a similar image first published by~\citet{BLIS5} and is used with permission.}
\label{fig:BLIS}
\end{figure}

\begin{figure}[htb!]
\centering
\begin{tabular}{l}
\textbf{Unpacked data can map poorly to cache associativity sets:} \\
\includegraphics[width=0.8\textwidth]{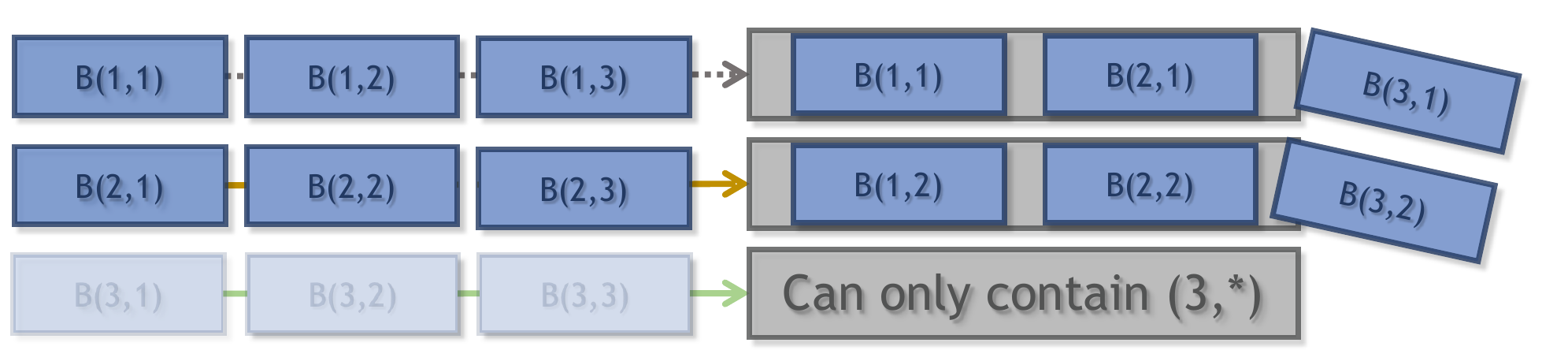} \\
\\
\textbf{Packed data maps to sets predictably and equitably:} \\
\includegraphics[width=0.8\textwidth]{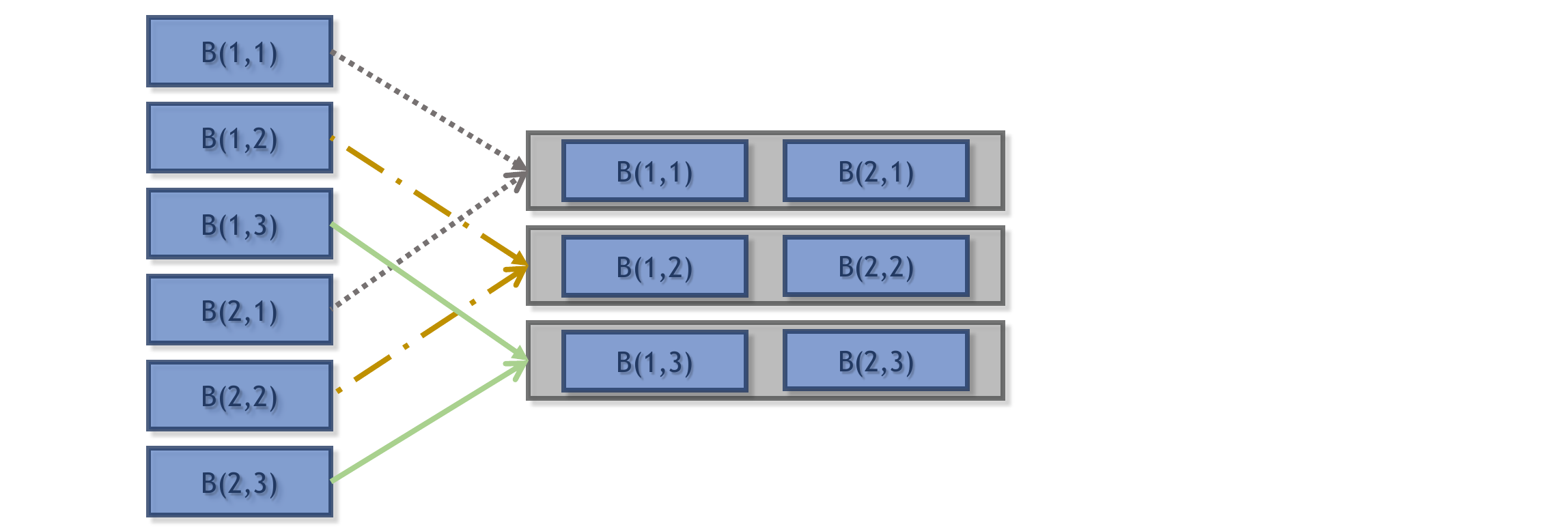}
\end{tabular}
\caption{
A schematic illustrating how the storage format of a micropanel in memory affects how cache lines map to the associativity sets of a cache. Top: Unpacked columns with arbitrary leading dimensions tend to cause more cache lines to be mapped to fewer sets of the cache, leading to inefficient use and unnecessary evictions. Bottom: The same submatrix stored contiguously causes cache lines to be mapped across all sets, leading to fewer evictions and, by proxy, fewer subsequent cache misses. A more rigorous analysis can be found in \citet{10.1145/2925987}.
}
\label{fig:pack2fit}
\end{figure}

Goto's algorithm structures a prototypical \gemm, $ C := A B + C $, where $A, B$, and $C$ are matrices of size $m \times k, k \times n$ and $m \times n$ respectively, as five loops around the update of a small submatrix  of $ C $ called the microtile, as illustrated in \cref{fig:BLIS}.
We only give the highlights here since the algorithm and this picture have been explained in many previously-published papers.
In this discussion and the figure, $ m_R $ and $ n_R $ denote register blocking sizes, while $ m_C $, $ n_C $, and $ k_C $ denote cache blocking parameters.
At the core is the \emph{microkernel}, which updates an $ m_R \times n_R $ \textit{microtile} of $ C $
by multiplying a $ m_R \times k_C $ \emph{micropanel} of $ A $ by a $ k_C \times n_R $ micropanel of $ B $.  On a typical architecture, the microtile of $ C $ is kept in registers while the micropanels of $ B $ and $ A $ are streamed from the L1 and L2 caches, respectively.
Blocks of $ A $ and row panels of $ B $ are rearranged (packed) at strategic points in the algorithm
to allow memory access with unit stride\footnote{This refers to loading consecutive memory addresses, which allows most CPUs to execute in the fewest number of cycles.} as well as to align micropanels from $A$ and $B$ so that they can fit into their designated levels of cache as indicated in \cref{fig:pack2fit}.
We refer to this rearranged storage as ``packed memory'' in contrast to ``unpacked memory'' used to store the original $A$ and $B$ matrices.

\subsection{BLIS's refactoring of Goto's algorithm}

The BLIS implementation of Goto's algorithm recognizes that as long as the microkernel is expressed with assembly code\footnote{This can take the form of so-called extended inline assembly code in addition to pure assembly code. Vector intrinsics \emph{may} also work, depending on the compiler and instruction set being emitted.}, high performance can be achieved even if all remaining parts of the algorithm above the microkernel (including packing) are written in C.
This reduces how much code must be customized for the \gemm\  operation.
It also allows other matrix-matrix operations supported by the BLAS (level-3 BLAS), such as Hermitian matrix-matrix multiplication (\textsc{hemm}), Hermitian rank-$k$ update (\textsc{herk}), triangular matrix-matrix multiplication (\textsc{trmm}), and triangular solve with multiple right-hand sides (\textsc{trsm}), to employ the same microkernel~\cite{BLIS1,BLIS2}.  
This contrasts with the original GotoBLAS implementation, inherited by OpenBLAS, where the two loops around the microkernel form what we call a \emph{macrokernel} that must be customized in assembly code for different \gemm-like operations~\cite{Goto2}.\footnote{When the assembly region extends to encompass the macrokernel, an OpenBLAS-like implementation may choose to maintain separate macrokernels for each operation \emph{or} insert conditional logic into a single macrokernel that allows the code to handle multiple similar operations (e.g. \textsc{syrk} and \textsc{herk}). The former case yields more regions of assembly code with only minor differences between them while the latter results in less assembly code that is nonetheless more difficult to decipher due to its embedded branching.}

An important detail is how to handle situations where the matrix dimensions are not whole multiples of $ m_R $, $ n_R $, and/or $ k_C $. These so-called ``fringe'' or edge cases.
During packing, BLIS pads micropanels with zeroes when a fringe case is encountered.
And since the microkernel author only needs to target microtiles of one size -- $ m_R \times n_R $ -- the job of writing and optimizing microkernels becomes much simpler.
The result is an easier-to-develop and easier-to-maintain code base at the cost of a minor decrease in performance for certain problem sizes~\cite{BLISPerformance}.

\subsection{SUP: Supporting small-ish matrices}
\label{sec:gemmsup}

More recently, projects and implementations like LIBXSMM~\cite{libxsmm} and BLASFEO~\cite{BLASFEO} have sought to improve matrix-matrix performance for small-sized problems.
Typically, these solutions either skip packing or require data to be pre-packed to exploit the fact that for small problems, matrices $A$ and $B$ have a chance to fit into the L2 (or even L1) cache in their entirety, thus avoiding the $\bigo(m k + n k)$ cost of packing.

BLIS's current approach to cases where at least one matrix dimension is small is referred to as Skinny/UnPacked (SUP).
It combines the following techniques:
\begin{itemize}
\item
It skips packing.
\item
Rather than isolating architecture specifics only in the microkernel, it employs a \emph{millikernel} that absorbs the first loop around the microkernel into the kernel primitive in an effort to reduce the frame stack cost (that is, memory overhead due to subroutine calls).
This millikernel is written in a manner similar to that of a corresponding microkernel (i.e., in assembly code).
\item
When a millikernel encounters a fringe case during its last iteration, it dispatches a helper microkernel that specializes in that size. Alternatively, a set of fringe cases may be called in sequence to emulate the net effect of a single, larger microkernel. For example, let us assume the size of the microtile is $6 \times 8$ and the millikernel encounters a fringe that is $ 5 \times 8$. This could result in a call to a single helper microkernel call that updates a $5 \times 8 $ microtile, or it could result in a call to a $3 \times 8$ helper microkernel followed by one that computes the remaining $2 \times 8$ part.
\item If a fringe case is encountered in the loop that \emph{surrounds} the millikernel (i.e., the 2nd loop around the microkernel), a more specialized millikernel is dispatched -- one that dispatches to a different subset of helper microkernels. Building on the previous example, if the millikernel is called on an $ m \times 5 $ submatrix, a helper millikernel that operates on at most 5 columns of $B$ is dispatched, which will then loop over a microkernel-like region of code targeting $6 \times 5$ and eventually, if needed at the fringe, dispatch helper microkernels whose $m_R$ values are less than 6.
\end{itemize}
While these changes to the conventional BLIS code path are conceptually simple, the examples above suggest that a nontrivial amount of code is required to support it.
Specifically, if a base-2 ``spanning'' set of kernels is employed, based on a $ 6\times 8$ microtile, the following millikernels and microkernels would be needed:
\newcommand{\mm}{\textsc{mm}}
\newcommand{\mh}{\textsc{m}\mu}
\newcommand{\hm}{\textsc{hm}}
\newcommand{\hh}{\textsc{h}\mu}
\begin{center}
\begin{tabular}{cccc}
$6\times8 (\mm)$ & $6\times4 (\hm)$ & $6\times2 (\hm)$ & $6\times1 (\hm)$ \\
$4\times8 (\mh)$ & $4\times4 (\hh)$ & $4\times2 (\hh)$ & $4\times1 (\hh)$ \\
$2\times8 (\mh)$ & $2\times4 (\hh)$ & $2\times2 (\hh)$ & $2\times1 (\hh)$ \\
$1\times8 (\mh)$ & $1\times4 (\hh)$ & $1\times2 (\hh)$ & $1\times1 (\hh)$
\end{tabular}
\end{center}
Here, $\mm$ denotes the main millikernel that is called from the 2nd loop, and $\mh$ denotes its helper microkernels. Similarly, the $\hm$ labels denote helper millikernels, each of which calls its own helper microkernels labeled by $\hh$.

\subsection{Combining the two code paths}

\begin{figure}[htbp]
\centering
\begin{tabular}{r@{\hspace{1ex}}c r@{\hspace{1ex}}c}
& DGEMM on Xeon E5-2690 & & DGEMM on AWS Graviton 3 \\
\rotatebox[origin=c]{90}{GFLOPS/sec} &
\raisebox{-0.5\height}{\includegraphics[width=0.435\textwidth]{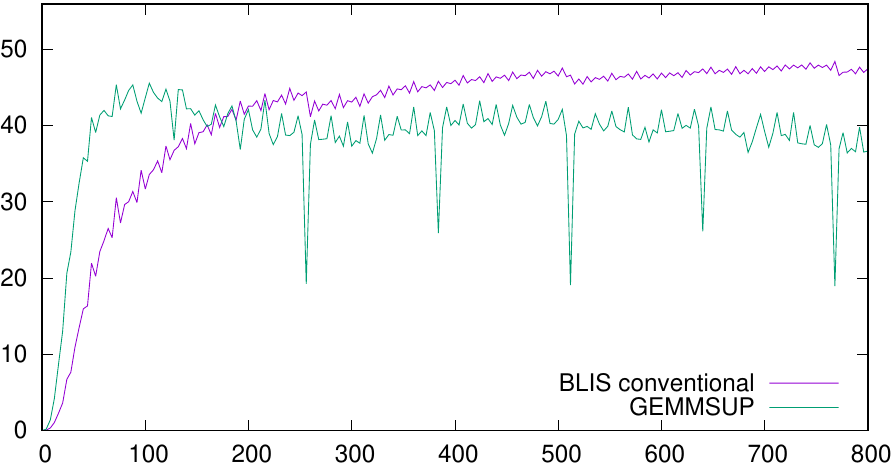}} & 
\rotatebox[origin=c]{90}{GFLOPS/sec} &
\raisebox{-0.5\height}{\includegraphics[width=0.435\textwidth]{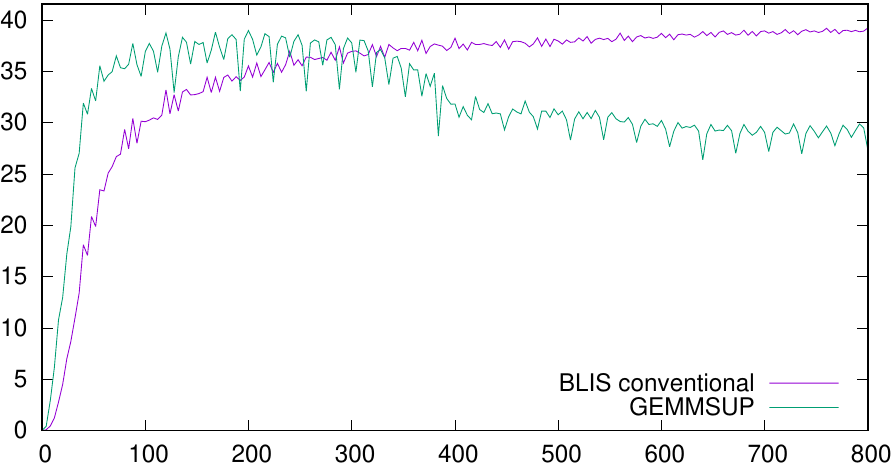}}
\\ & $ m = n = k $ & & $ m = n = k $
\end{tabular}
\caption{Performance of: BLIS's conventional \gemm\ and \gemm\ implemented with the SUP approach (\gemmsup) on the Intel Xeon E5-2690 (left) and AWS Graviton 3 with \armsve\ (right) architectures. Note the different cross-over points for each system.}
\label{fig:perf_crossover}
\end{figure}

\Cref{fig:perf_crossover} reports performance of the conventional BLIS \gemm\ and the \gemm\ implemented via the SUP approach (\gemmsup) on two different architectures.
Clearly, SUP outperforms the conventional algorithm for small problem sizes.
However, as the problem size becomes large, data would unavoidably spill out of the cache, degrading SUP's performance and creating periodic spikes down on the Xeon E5-2690 processor.
Importantly, to roll out solutions to a matrix library, the cross-over point must be determined and encoded.
The heuristic for this is complicated by the fact that it is not only a function of the problem size but also of the row and column strides for matrices $ A $ and $ B $.
Additionally, there may be an architecture-dependent range of sizes where both algorithms suffer performance degradation (e.g., around $ m = n = k = 200 $ on the Xeon E5-2690).
In such a region, unaligned
SUP may spill data out of the cache, while the $\bigo(m k + n k )$ packing cost for the conventional algorithm is non-negligible.
 We will soon show that fusing the packing with computation can address these issues.

\section{A Unified Approach}

It would appear that the packing process inherently imposes unreasonable overhead for small- to medium-sized matrix cases.  We now discuss how that is not always so.

\subsection{Interleaving of packing and computing}
\label{sec:interleaving}

\begin{algorithm}[htbp]
\caption{Algorithm with interleaved packing and computation.}
\label{fig:simplemod}
\For{ $\cdots$ (the \nth{5} and \nth{4} loops around the microkernel proceed as before but without packing) $\cdots$ }{
\For{ first iteration of \nth{3} loop around the microkernel }{
    \For{ the first iteration of \nth{2} loop around the microkernel }{
        Pack the first micropanel of $ B_{p,j} \rightarrow \widetilde B_{p,j} $\;
        \For{ each iteration of \nth{1} loop around the microkernel }{
            Pack current micropanel of $ A_{i,p}  \rightarrow \widetilde A_{i,p} $\;
            Call the microkernel with packed micropanels $ \widetilde A_{i,p} $ and $ \widetilde B_{p,j} $\;
        }
        Upon completion, $ A_{i,p} $ is left packed in $ \widetilde A_{i,p}$\;
    }
    \For{ remaining iterations of \nth{2} loop around microkernel }{
        Pack current micropanel of $ B_{p,j} \rightarrow \widetilde B_{p,j} $\;
        \For{ each iteration of \nth{1} loop around the microkernel }{
            Call the microkernel with packed micropanels $ \widetilde A_{i,p} $ and $ \widetilde B_{p,j} $\;
        }
    }
    Upon completion, $ B_{p,j} $ is left packed in $ \widetilde B_{p,j}$\;
}
\For{ remaining iterations of \nth{3} loop around microkernel }{
    \For{ the first iteration of \nth{2} loop around the microkernel }{
        \For{ each iteration of \nth{1} loop around the microkernel }{
            Pack current micropanel of $ A_{i,p}  \rightarrow \widetilde A_{i,p} $\;
            Call the microkernel with packed micropanels $ \widetilde A_{i,p} $ and $ \widetilde B_{p,j} $\;
        }
        Upon completion, $ A_{i,p} $ is left packed in $ \widetilde A_{i,p}$\;
    }
    \For{ remaining iterations of \nth{2} loop around microkernel }{
        \For{ each iteration of \nth{1} loop around the microkernel }{
            Call the microkernel with packed micropanels $ \widetilde A_{i,p} $ and $ \widetilde B_{p,j} $\;
        }
    }
}}
\end{algorithm}

The basic idea behind unifying is simple:  Modify Goto's algorithm so that elements of the micropanels of $ A $ and $ B $ are packed \emph{just before} their first use by a call to the microkernel.  More precisely, consider that Goto's algorithm has proceeded to the first iteration of the 4th loop, in which a column-panel of $ A $ is multiplied by a row-panel of $ B $, except let us assume that neither is packed:
\begin{center}
\includegraphics[width=0.55\textwidth]{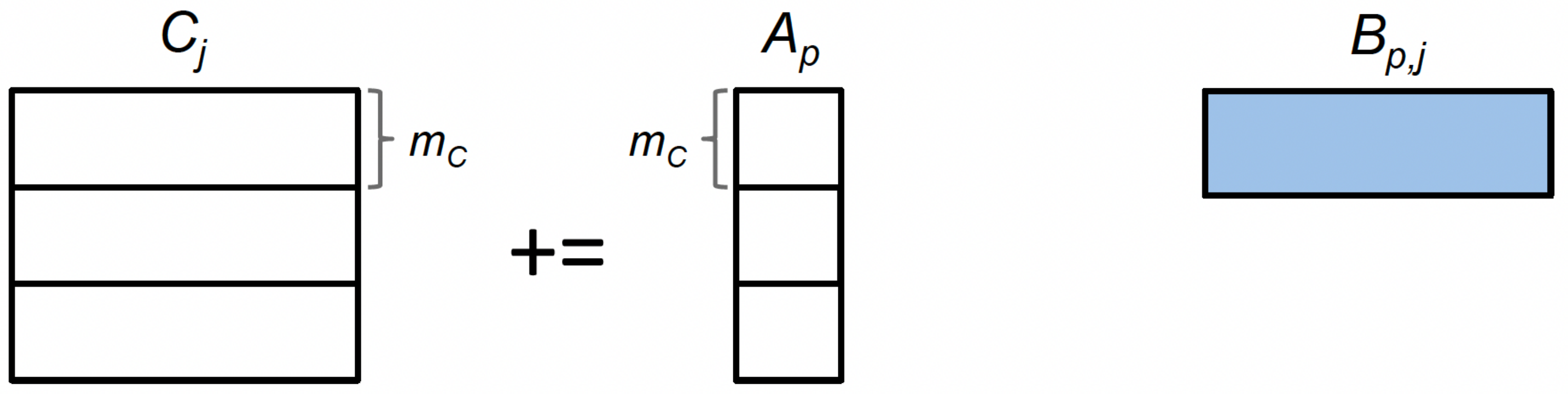}.
\end{center}
This operation can be implemented by modifying Goto's algorithm as described in \cref{fig:simplemod}.
It allows the microkernel from the  conventional BLIS implementation to be used without modification.

The benefit of this interleaving is that after packing, micropanels of $ \widetilde B_{p,j} $ and $ \widetilde A_{i,p} $ are still in the L1 cache when used for the first time by the microkernel, while their eventual migration back to the L3 and L2 caches, respectively, are (likely) masked by computation. 
This can be expected to improve performance in general, but in particular for small matrices.

\subsection{FIP: Fused packing in the microkernel}
\label{subsec:fused_packing}

While the simple solution mentioned in the previous subsection should reduce the net execution time of the microkernel for the first time a packed micropanel is involved in computation, packed data still moves from registers to the L1 cache and back. 
This ``reflowing'' of data adds cycles to the total execution time, and can sometimes trigger a performance penalty for \emph{read-after-write} (RAW) data hazards especially on some \ia\ architectures.

Our solution is to fuse individual packing instructions into the microkernel itself.
We call this technique ``fused-in packing'' (FIP).

Conceptually, this is similar to the previous approach, except that the unit of data around which the packing and computation are interleaved is reduced from whole micropanels of $A$ and $B$ to a mere handful of elements.
This has at least two benefits:  
(1) Data that is loaded into registers in the normal course of packing is reused immediately for useful computation;
(2) The cost of loading data from the cache into registers is partially hidden by the computation on previously-loaded values.

Implementing FIP requires instantiating four cases of the microkernel:
\begin{itemize}
\item
The conventional microkernel, where microtiles from both sides are already packed.
\item
A microkernel where $ \widetilde A_{i,p} $ is already packed but $ B_{p,j} $ is unpacked (and thus needs to be packed).
\item
A microkernel where $ \widetilde B_{p,j} $ is already packed but $ A_{i,p} $ is unpacked (and thus needs to be packed).
\item
A microkernel where both $ A_{i,p} $ and $ B_{p,j} $ are unpacked.
\end{itemize}
Each of these might be encountered while computing $ C := AB + C $ depending on the sizes of the matrices.

\subsection{Options in fused packings}
\label{subsec:fused_packing_opt}

There are some situations in which there is little benefit from packing $ B_{p,j} $ and/or $ A_{i,p} $.
Consider a \gemm\ call where $ C $ is $ m \times n $, $ A $ is $ m \times k $, and $ B $ is $ k \times n $.
If $ n \leq n_R $, then the 2nd loop around the microkernel is only executed once, and hence packing $ A_{i,p} $ would yield no benefit.
And if $ m \leq m_R $ then $ B_{p,j} $ is similarly not reused.
The resulting spectrum of options can be summarized as
\begin{center}
\begin{tabular}{| c || c | c |} \hline
& $n \leq n_R$ & $n > n_R$ \\ \hhline{|=||=|=|}
$ m \leq m_R $ & no packing  & pack $ A_{i,p} $  \\ \hhline{|-||-|-|}
$ m > m_R $ & pack $ B_{p,j} $ & pack $ A_{i,p} $ and $ B_{p,j} $ \\ \hline
\end{tabular}
\end{center}
This provides a decent heuristic from which further tuning may be explored.

Another observation concerns the packing of $ A_{i,p} $.  Let us assume that $ A $ is stored in column-major order.
If 
\[
{\rm csb}(A) \times k_C 
\leq 
\rm{(L2~cache~size~in~bytes)},
\]
where $ {\rm csb}(A) $ equals the stride in bytes between elements in a row of $ A $, 
then the unpacked storage of $ A_p $ will never cause an L2 cache spill -- that is, elements of $ A_p $ will not be evicted from the L2 cache by other elements.
This constitutes another case where we can skip the packing of $ A $.
If $ A $ is stored in row-major order, the condition becomes:
\[
m_C \times {\rm rsb}(A) 
\leq 
\rm{(L2~cache~size~in~bytes)},
\]
where $ {\rm rsb}(A) $ equals the stride in bytes between elements in a column of $ A $.
The details behind these inequalities go beyond the scope of this paper and require an understanding of the results by~\citet{10.1145/2925987}.

\begin{figure}[htbp]
\centering
\begin{tabular}{r@{\hspace{1ex}}c r@{\hspace{1ex}}c}
& DGEMM on Xeon E5-2690 & & DGEMM on AWS Graviton 3 \\
\rotatebox[origin=c]{90}{GFLOPS/sec} &
\raisebox{-0.5\height}{\includegraphics[width=0.435\textwidth]{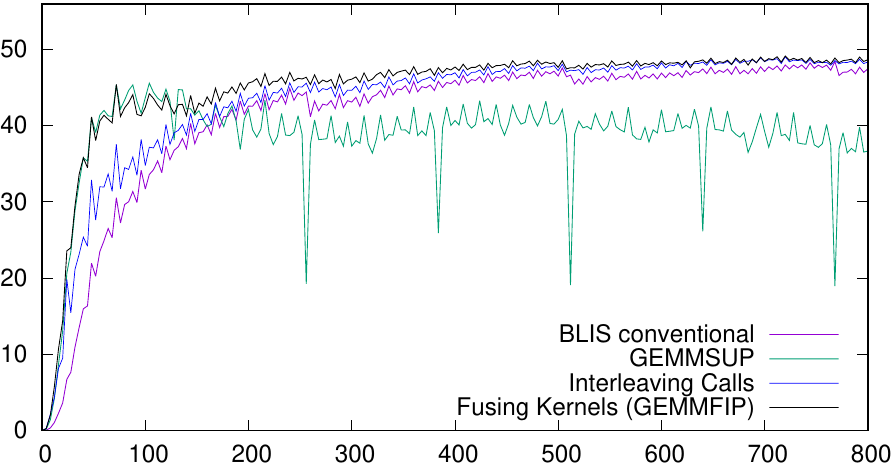}} &
\rotatebox[origin=c]{90}{GFLOPS/sec} &
\raisebox{-0.5\height}{\includegraphics[width=0.435\textwidth]{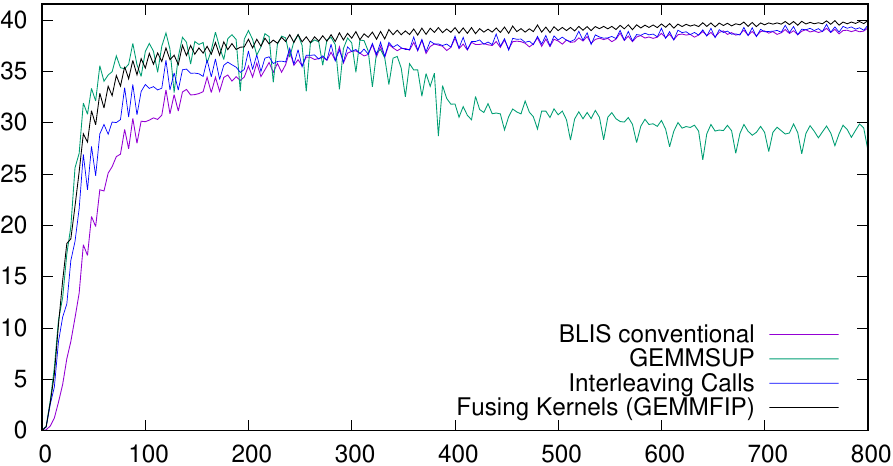}}
\\ & $ m = n = k $ & & $ m = n = k $
\end{tabular}
\caption{Performance improvements observed on the Intel\textregistered\ Xeon E5-2690 (top) and the AWS Graviton 3 (bottom) processors from interleaving packing and computing and optionally fusing their kernels (FIP/\gemmfip).}
\label{fig:simplemod_perf}
\end{figure}

\Cref{fig:simplemod_perf} illustrates how interleaving packing and computing as \cref{fig:simplemod} benefits the performance of \gemm\  and how deploying the FIP technique further improves it.
Additional experiments are available in \cref{sec:performance}.

\subsection{Coding effort}

\begin{algorithm}[htb!]
\caption{A code sample for multi-instantiating SUP-based kernels with packing instructions conditionally (depending on the values of macro arguments \texttt{pa} and \texttt{pb}) fused in to handle all the four cases required by the FIP approach.}
\label{alg:multi_instantiate}
\begin{lstlisting}[language=C]
#define cond_inst_false(_1)
#define cond_inst_true(_1) _1

#define kernel_def(pa, pb) \
\
void fused_kernel_## pa ##_## pb ( /* func params */ ) \
{ \
  __asm__ volatile \
  ( \
    /* ... */ \
\
    /* Multiply and accumulate: ymm0 holds A, ymm2 holds B and ymm4 holds C fractions */ \
    "vfmadd231pd %ymm0, %ymm2, $ymm4 \n\t" \
\
    /* Optionally store ymm2 to the packing space indicated by rdx. */ \
    cond_inst_## pb ( "vmovapd %ymm2, 16(%rdx) \n\t" ) \
\
    /* ... */ \
  ) \
}

kernel_def(true, true)
kernel_def(true, false)
kernel_def(false, true)
kernel_def(false, false)
\end{lstlisting}
\end{algorithm}

We implemented FIP kernels based on the microkernel portion of the SUP code path in BLIS introduced in \cref{sec:gemmsup}.
Though \cref{subsec:fused_packing,subsec:fused_packing_opt} imply that the approach requires handling all four cases within one set of microkernels, most of the redundancies between the four cases can be tamed with some careful refactoring.
BLIS kernels are written with inlined assembly code in C.
Meanwhile, fused packing merely requires data from registers to be moved to memory with hard-codable strides.
These two facts allowed us to leverage the C preprocessor to expand a single code template into the four specializations in a fashion shown in \cref{alg:multi_instantiate}.

\subsection{Multithreading}
\label{subsec:threading}

\begin{figure}[htb!]
\centering
\includegraphics[width=0.875\textwidth]{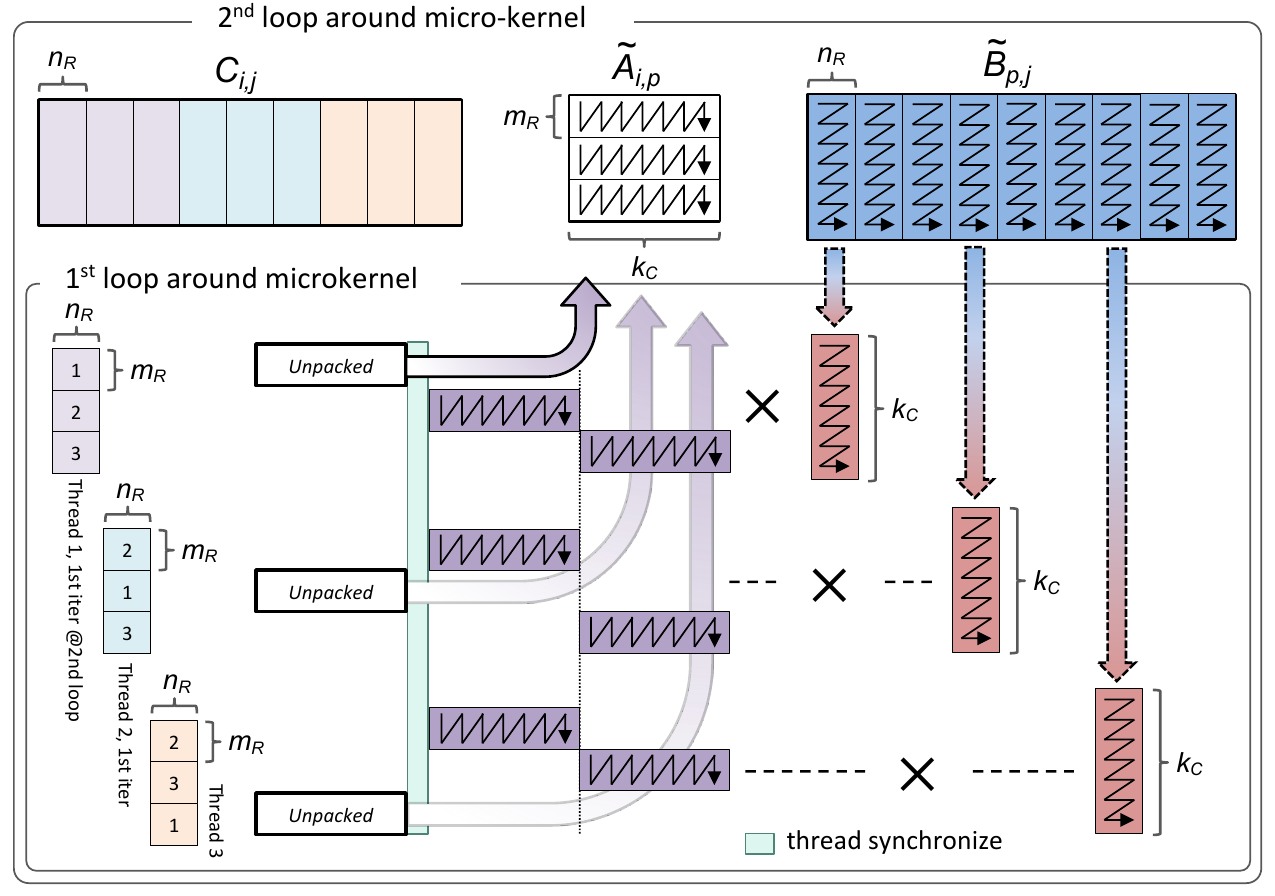}
\caption{An illustration of how multithreading can be added to our FIP approach.
Each thread starts from a different unpacked micropanel of $A_p$ and packs the micropanel to its designated space in $\widetilde A_{i,p}$ while performing the corresponding update of a microtile of $C: C_{i,j} $. Once all threads have finished packing their $\left.\frac{m_C}{m_R}\middle/ n_\mathrm{thr} \right.$ micropanels, \emph{one} synchronization occurs so they can use all micropanels in $\widetilde A_p$. 
This collaborative packing and computing only happen when each thread is working on its first $B_{p,j}$ micropanel within the \nth{2} loop around the microkernel.}
\label{fig:gemmhot_mt}
\end{figure}

We now take a brief look at the multithreading potential of the FIP technique, which is left for future research on this approach.

BLIS's refactoring of Goto's algorithm conveniently exposes five loops coded in C99 in which multithreading can be introduced~\cite{BLIS3}. In practice, parallelism is usually gained
in the 2nd loop around the microkernel
to let multiple threads operate on the same $m_C \times k_C$ tile from $A$ or
in the  3rd loop around the microkernel
to let multiple threads operate on the same $k_C \times n_C$ panel from $B$.
On a multi-core chip with non-uniform memory access (NUMA) architecture, cores usually share L2 caches within each NUMA, making it reasonable to parallelize over the 2nd loop within and the 3rd loop across NUMA nodes.

With this insight, if we want to apply our insight to the consumption of tiles from $A$, it seems appropriate to let each thread work on the unpacked memory first and collaboratively write the microtiles they have loaded to the packing space for later consumption.
Since threads begin referring to the packed data only from their second iteration of the 2nd loop around the microkernel, for $A_p$ it is only required that all threads working on the same $m_C \times k_C$ tile from $A$ synchronize \emph{once} after finishing their first iteration there. This synchronization cost is identical to collaboratively packing everything from $A$ beforehand.

Furthermore, if one still wants to ensure each tile from unpacked memory is accessed only once, we can even change the order when each iteration of the first loop around the microkernel gets executed. Letting each thread enter the microkernel at a different microtile of $C$ allows the requested $m_C \times k_C$ to become readily packed after each thread has finished their first $\left. \frac{m_C}{m_R} \middle/ n_\text{thr} \right.$ iterations, as is illustrated in \cref{fig:gemmhot_mt}. Here $n_\text{thr}$ is denoted as the number of threads working on the same $m_C \times k_C$ tile from $A$ and $m_C$ is supposed to be multiples of $m_R n_\text{thr}$.

On the $B$ side, in each iteration within the 2nd loop around the microkernel, the microtile from $B$ is reused from the thread-private L1 cache and no distinct gain in performance can be expected from reusing micropanels of $B$ from the L3 cache.  In addition, thread synchronization across NUMA sockets is likely to cause nontrivial overhead. 
This suggests that it is better to let each thread pack its relevant panels into a separate space despite the modest storage redundancy. The threads then flow data from their private L1 cache to the registers without interfering with each other.

\section{Performance results}
\label{sec:performance}

We now illustrate the benefits of the discussed technique on a broad range of architectures, for a single core.  

\subsection{Experimental setup}
\label{subsec:setup}

We implemented \gemm\  with the FIP approach (\textsc{gemmfip}) based on BLIS (Release 0.9.0).
Our implementation can be integrated back into BLIS as an alternative backend for \gemm\  through its \emph{sandbox} interface.
The macrokernel structure of our implementation is essentially the same as BLIS's refactoring of Goto's algorithm in \cref{fig:BLIS} with a few additional lines to handle kernel selection for the four cases mentioned in \cref{subsec:fused_packing}.

A non-trivial modification here is for \ia\  kernels. As we have already mentioned in \cref{sec:gemmsup}, to minimize costs associated with pushing to the frame stack, BLIS's SUP kernels include the first loop around the microkernel (i.~e.~the millikernel) in the assembly region of the code.
In our implementations, this was also incorporated for \ia -based architectures since they have a limited number of general-purpose registers compared to \arma \textregistered, Power ISA, and RISC-V\textregistered, making it harder for compilers to transition between microkernel calls without interfering with memory consistency or causing frame stack RAW data hazards.

We developed microkernels to support double-precision \gemmfip\  on 3 architectures: Intel AVX2, \arm\  and \armsve.
Performance experiments were performed on five different processors: 
Intel\textregistered\  Xeon E5-2690
(3500 MHz, compiler: GCC 10.2),
AMD Epyc\texttrademark\  7R32 with \ia\  AVX2 architecture
(3300 MHz, compiler: GCC 11.3),
AWS Graviton 2 with \arm\  (2500 MHz, compiler: Clang 14.0), AWS Graviton 3 with \armsve\  (2600 MHz, compiler: Clang 14.0), 
and Apple M2 with \arm\  (3490 MHz, compiler: Apple Clang 14.0).

All our experiments used a single core.
The uppermost depicted $y$-axis value represents the single-core theoretical peak performance for each processor tested.

\subsection{Evaluation}

\input PerformanceGraphsIntelAMD
\input PerformanceGraphsARM

In \cref{fig:perf_gemm_ARM,fig:perf_gemm_IntelAMD} (top), we report the performance of the conventional BLIS code path, its SUP code path, and \gemmfip.
In the top graphs, the leading dimension (\lda) -- that is, the stride between logically adjacent elements within in a single row -- equals the (row) dimension of the matrices.
In \cref{fig:perf_gemm_ARM,fig:perf_gemm_IntelAMD} (center-top), the leading dimension equals $ 2000 $, which can negatively affect performance, particularly when computing on unpacked data.

We see that the curve for \gemmfip\ uniformly matches or outperforms the conventional BLIS and SUP paths while smoothly bridging the ``medium-sized'' region where both underperform.
This result supports the claim that the new method can provide a unified approach that yields high performance across the range of problem sizes without having to deploy heuristics to determine a crossover point between the paths.

\subsection{Comparison with other BLAS implementations}
\label{sec:more_perf}

We now compare how our implementation of \gemmsup\  performs against other matrix libraries, including BLIS itself, OpenBLAS, and numerous vendor-specific BLAS implementations.

On the \ia\  side, Intel's  Math Kernel Library (MKL) is akin to the ``gold standard.''  Since MKL is a closed-source implementation, it is difficult to discern what allows that code to perform so well.
In \cref{fig:perf_gemm_IntelAMD} (center-bottom and bottom), our implementation handily outperforms OpenBLAS and matches or surpasses BLIS, especially for medium and even large problem sizes.
On AMD's Zen\texttrademark -microarchitecture-based processors, BLIS is integrated into the AMD Optimizing CPU Libraries (AOCL) with vendor-side tunings.
AMD also has an open-source fork of BLIS known as AMD BLIS, which we have built and tested apart from AOCL. On that processor, \gemmfip\  essentially matches AOCL's performance (with the exception of a narrow range from about $150$ to $250$ when $\mathrm{LDim} = m$) while handily outperforming BLIS, AMD BLIS, and OpenBLAS.
Meanwhile, MKL yields mediocre (and inconsistent) performance on the 7R32.
We imagine that Intel made a conscious decision to throttle the performance of MKL when running on Epyc hardware in hopes of discouraging their users from switching to their competitor's products.

For \arma\  processors, \arma\  Performance Libraries (\arma PL) provides a state-of-the-art matrix library solution. Comparisons of \gemmfip\  against \arma PL, OpenBLAS, and BLIS are plotted in \cref{fig:perf_gemm_ARM} (center-bottom and bottom).
On the AWS Graviton 2, \gemmfip\  outperforms \arma PL and slightly underperforms OpenBLAS. This defect is presumably because our microkernels for unpacked memories still leave room for improvement on this specific microarchitecture, as can be deduced from the fact that \gemmsup\ performance using a similar microkernel lags far behind.
For Apple's M2 processor, its Accelerate framework was, upon closer inspection, found to be using a hidden co-processor~\cite{AppleAMX2} shared by the whole chip instead of the \arm\ pipelines, and the peak performance turned out to be around 380 GFLOPS/sec for $m = n = k = \left(\lda \textrm{ of } A, B, \textrm{and } C \right) \approx 800$, regardless of threading options given. This difference on the hardware side makes it impossible to measure an isolated single-core throughput. Therefore, the Accelerate framework is producing curves in \cref{fig:perf_gemm_ARM} that exceed the $y$-axis limits of the graphs.
Finally, on the \armsve\ architecture of the AWS Graviton 3 processor, \gemmfip\  yields the highest throughput and the best consistency among all tested libraries, demonstrating its value in this relatively nascent architecture.

\section{Conclusion}
\label{sec:conclusion}

We have described and demonstrated the benefits of reorganizing the memory access patterns associated with the Goto algorithm, as refactored by BLIS, so that small- and large-size problems are handled within a unified code path.
This novel approach integrates the packing instructions for $A$ and/or $B$ into the microkernel, where they may be templatized and conditionally activated at compile-time.
Fusing the packing and computation in this manner creates a spectrum of algorithms, where the SUP approach (no packing) and the Goto algorithm (full ``classic'' packing) each appear as special cases.
A runtime logic then tracks whether there is a reuse of $A_{i,p}$ and $B_{p,j}$ micropanels to determine which variant should be deployed to specific problem sizes.
Since the result yields uniformly high performance insensitive to the switching points in that runtime logic, our unified FIP approach appears to largely obviate the need for empirical crossover points or other runtime heuristics.

We have demonstrated the method's benefits by creating implementations for multiple architectures.
Only on the Intel Xeon E5-2690 processor does another implementation -- Intel's MKL -- outperform FIP.
Given MKL's reputation for achieving extremely high performance, we believe this clearly illustrates the importance of the work.
We expect these techniques to be adopted by BLIS and other libraries in the near future, thus further positively impacting the user community, regardless of which library they use.

The results in this paper suggest future work that will have an extended impact.
The most obvious is that the techniques can be applied to other precisions and other matrix-matrix multiplications (level-3 BLAS).
In addition, a body of papers shows how BLIS's refactoring of Goto's algorithm can be used to attain high performance and/or reduce the development effort for various \gemm-like operations:
\begin{itemize}
\item
In the work by~\citet{BLIS1m}, BLIS's \textsc{1m} method leverages the real-domain microkernel to implement complex-domain matrix-matrix multiplication operations by cleverly encoding the definition of complex scalar multiplication within the packing stage of Goto's algorithm.   
\item
The work by~\citet{Huang:2016:SAR:3014904.3014983} shows how Strassen's algorithm can already attain high performance for rank-$k$ updates and relatively small matrices.  
\item
\citet{Chenhan:SC15} give a high-performing implementation for solving the $k$-nearest neighbor problem by fusing computation into Goto's algorithm.
\item
In the work by \citet{MOMMS}, it is reasoned and demonstrated that Goto's algorithm is but one algorithm in a larger family of algorithms, the Multilevel Optimized Matrix-matrix Multiplication Sandbox (MOMMS) family.  The idea is that as the speed ratio between CPU arithmetics and memory access becomes worse in the future, blocking for caches must be modified.
\item
\citet{tblis_sisc} uses BLIS's refactoring of Goto's algorithm to instead implement tensor contractions, yielding TBLIS, by recognizing that rearrangement of data to cast tensor contraction as a matrix multiplication can be incorporated into packing.
\end{itemize}
Insights demonstrated in this paper can potentially accord base-performance benefits for all these algorithms.

Finally, if the strategy in \cref{subsec:threading} could bring our insight's unifying effect to the multithreading regime, perhaps it could extend to and yield speedup on GPUs whose intermediate-level memory operates in a cache-like way or as a register-controlled scratchpad memory (SPM)\footnote{For GPUs whose L1-level storage is an SPM with controllable direct memory access (DMA), it is expected that packing via DMA will provide better performance.}
as well.

\section{Code Availability}

Our code is developed under a fork of BLIS available at \url{https://github.com/xrq-phys/blis}. It may be enabled as an ``sandbox'' that is optionally integrated into the library by configuring BLIS as:
\begin{center}
    \tt
    ./configure -s gemmfip -t none x86\_64
\end{center}
for \ia\ microarchitectures or:
\begin{center}
    \tt
    ./configure -s gemmfip -t none arm64
\end{center}
for \arma\ hardware.

\section*{Acknowledgements}
We thank Prof.~S.~Todo for supervising RuQing Xu's research and providing access to various architectures.  We also thank members of the BLIS community for their input.

RuQing Xu is funded by The University of Tokyo's GSGC scholarship.  The researchers at The University of Texas at Austin are funded in part by the National Science Foundation (Award CSSI-2003921) and gifts from AMD, Arm, and Oracle.  

{\it Any opinions, findings, and conclusions or recommendations expressed in this material are those of the author(s) and do not necessarily reflect the views of the National Science Foundation.}


\printbibliography

\end{document}

%% file: PerformanceGraphsIntelAMD.tex
\begin{figure*}[p]
\newcommand{\graphwidth}{0.31\textwidth}

\centering
\begin{tabular}{r @{\hspace{1ex}} c @{\hspace{1mm}} c}
&
Xeon E5-2690 &
Epyc 7R32
\\
\rotatebox[origin=c]{90}{GFLOPS/sec w/ $\lda = m$} &
\raisebox{-0.5\height}{\includegraphics[width=\graphwidth]{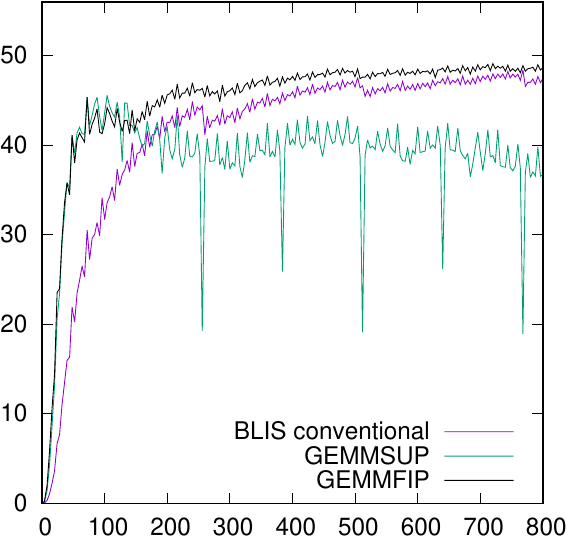}} &
\raisebox{-0.5\height}{\includegraphics[width=\graphwidth]{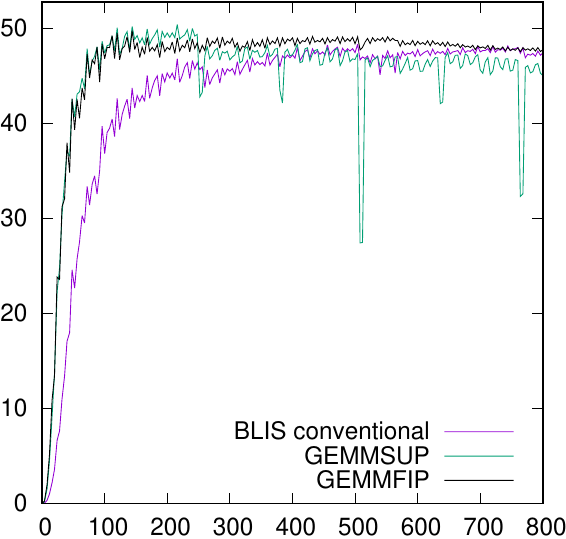}}
\vspace{1.5ex} \\
\rotatebox[origin=c]{90}{GFLOPS/sec w/ $\lda=2000$} &
\raisebox{-0.5\height}{\includegraphics[width=\graphwidth]{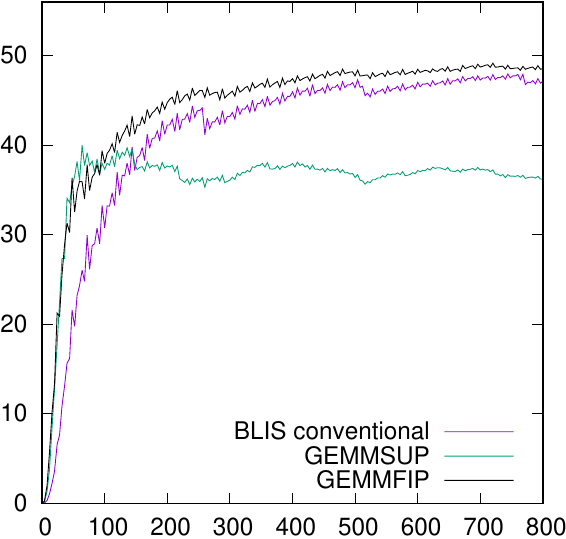}} &
\raisebox{-0.5\height}{\includegraphics[width=\graphwidth]{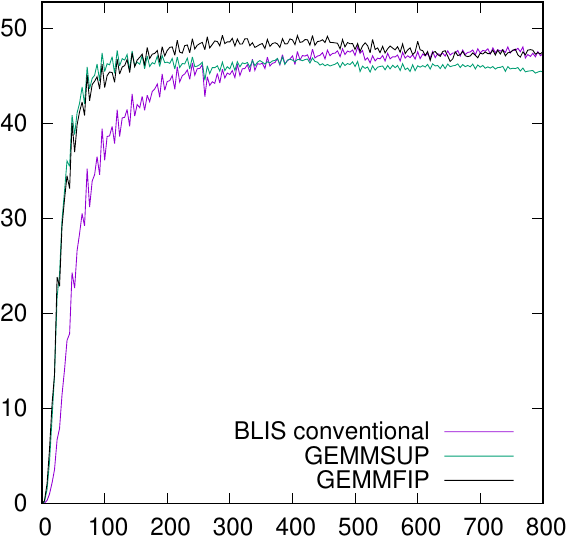}}
\vspace{1.5ex} \\
\rotatebox[origin=c]{90}{GFLOPS/sec w/ $\lda=m$} &
\raisebox{-0.5\height}{\includegraphics[width=\graphwidth]{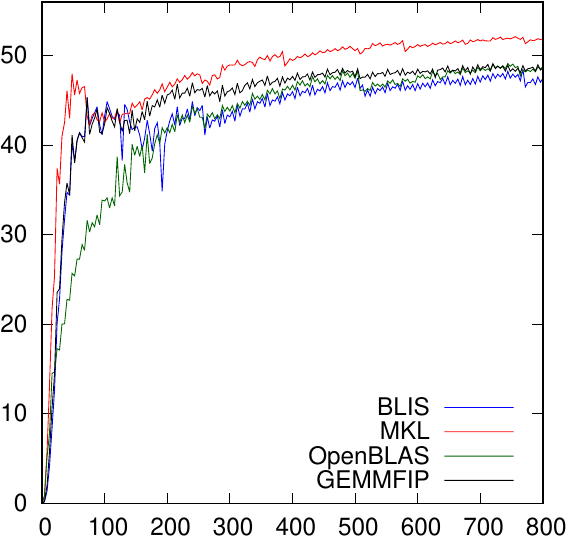}} &
\raisebox{-0.5\height}{\includegraphics[width=\graphwidth]{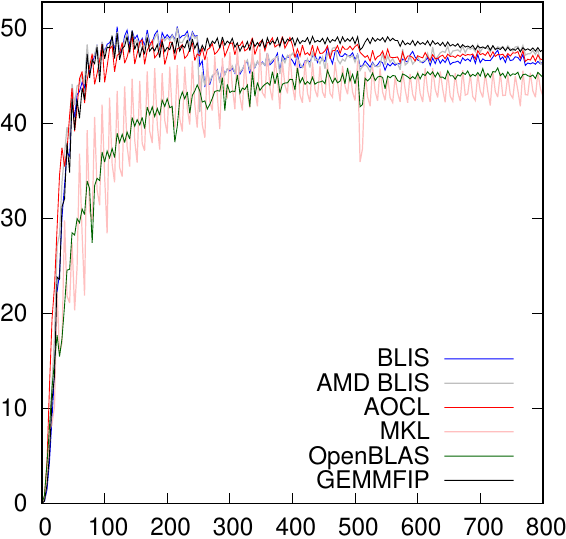}}
\vspace{1.5ex} \\
\rotatebox[origin=c]{90}{GFLOPS/sec w/ $\lda=2000$} &
\raisebox{-0.5\height}{\includegraphics[width=\graphwidth]{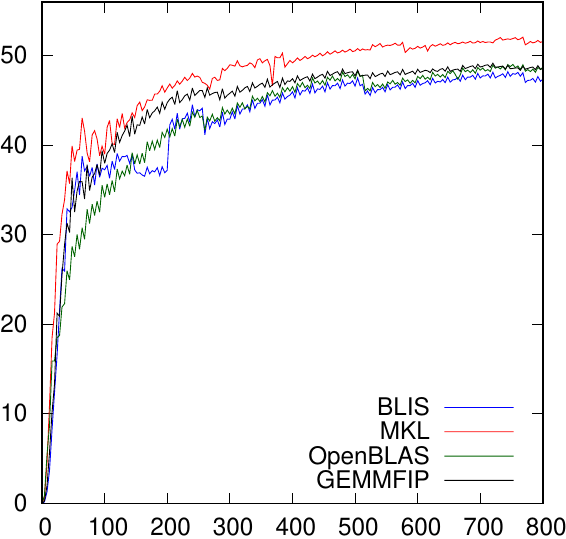}} &
\raisebox{-0.5\height}{\includegraphics[width=\graphwidth]{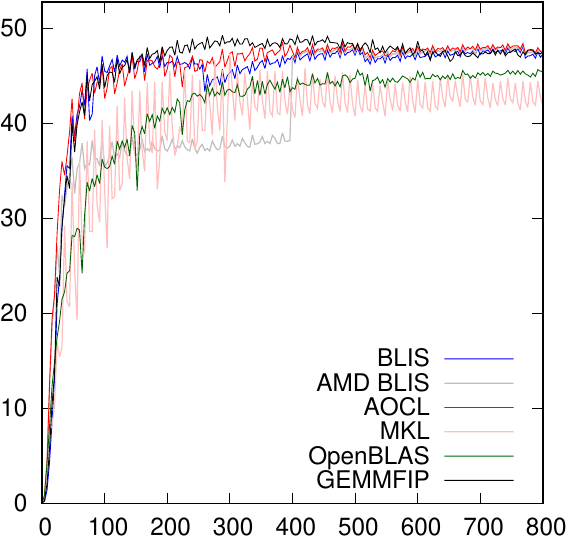}}
\vspace{1.5ex} \\
& $m = n = k$ & $m = n = k$
\end{tabular}
\caption{Performance on the Intel\textregistered\  Xeon E5-2690  (left) and AMD Epyc\texttrademark\  7R32 with \ia\  AVX2 (right) architectures.
}
\label{fig:perf_gemm_IntelAMD}
\end{figure*}

%% file: PerformanceGraphsARM.tex
\begin{figure*}[p]
\newcommand{\graphwidth}{0.31\textwidth}

\centering
\begin{tabular}{r @{\hspace{.6ex}} c @{\hspace{.6ex}} c @{\hspace{.6ex}} c}
&
AWS Graviton 2 &
AWS Graviton 3 &
Apple M2
\\
\rotatebox[origin=c]{90}{GFLOPS/sec w/ $\lda = m$} &
\raisebox{-0.49\height}{\includegraphics[width=\graphwidth]{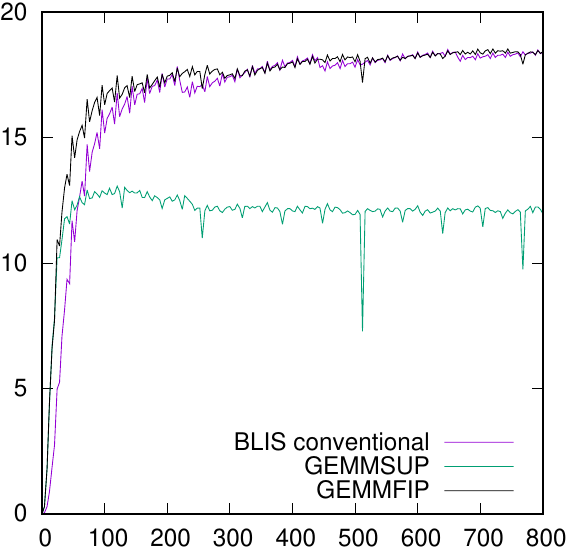}} &
\raisebox{-0.50\height}{\includegraphics[width=\graphwidth]{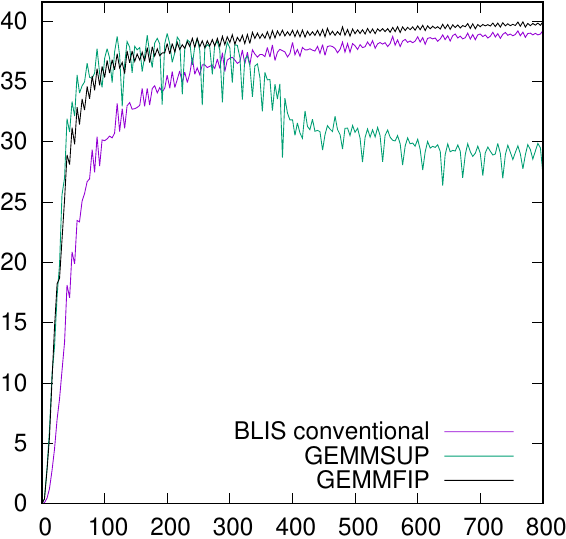}} &
\raisebox{-0.50\height}{\includegraphics[width=\graphwidth]{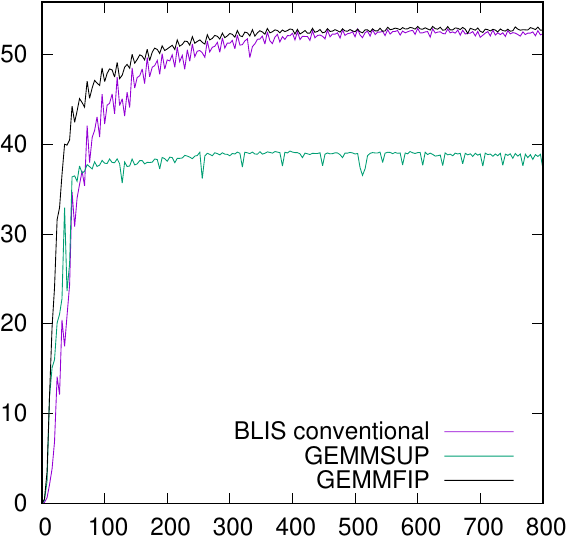}}
\vspace{1.5ex} \\
\rotatebox[origin=c]{90}{GFLOPS/sec w/ $\lda = 2000$} &
\raisebox{-0.49\height}{\includegraphics[width=\graphwidth]{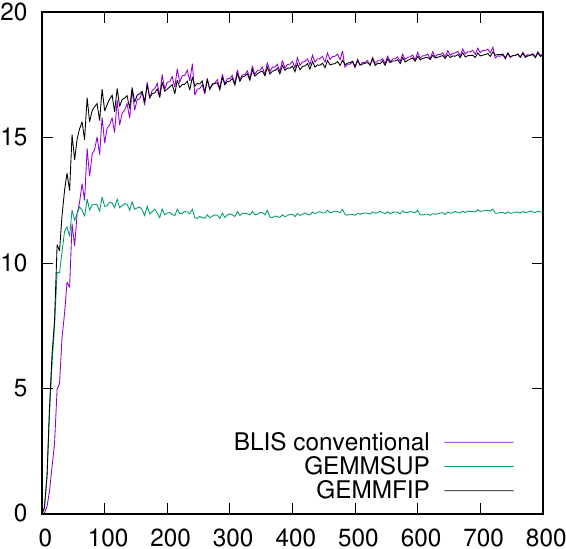}} &
\raisebox{-0.50\height}{\includegraphics[width=\graphwidth]{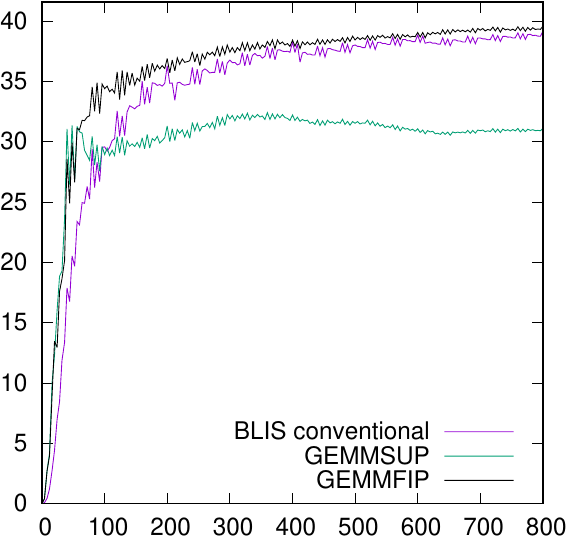}} &
\raisebox{-0.50\height}{\includegraphics[width=\graphwidth]{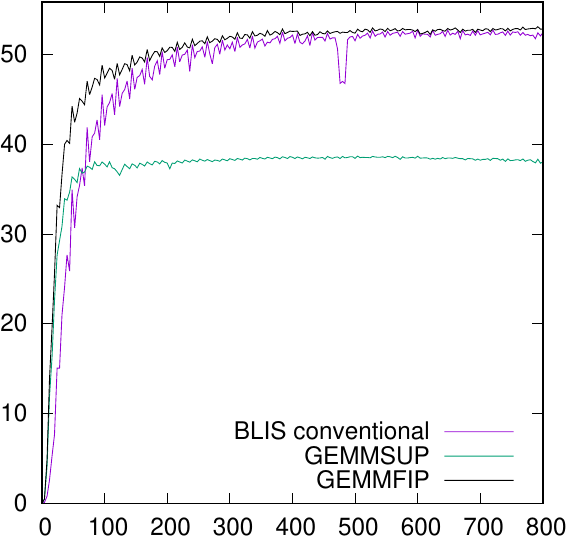}}
\vspace{1.5ex} \\
\rotatebox[origin=c]{90}{GFLOPS/sec w/ $\lda = m$} &
\raisebox{-0.49\height}{\includegraphics[width=\graphwidth]{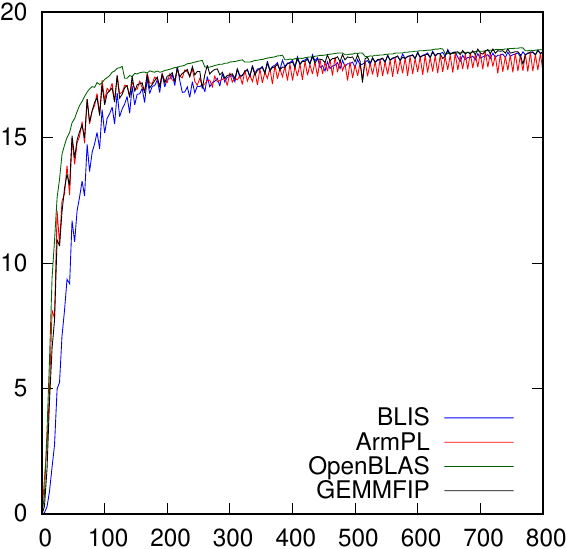}} &
\raisebox{-0.50\height}{\includegraphics[width=\graphwidth]{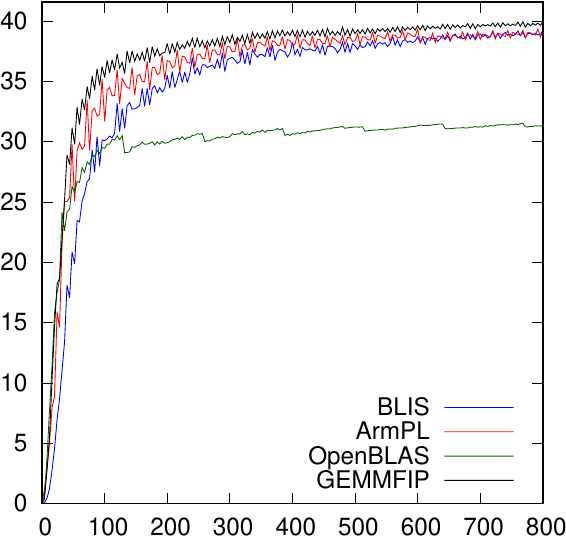}} &
\raisebox{-0.50\height}{\includegraphics[width=\graphwidth]{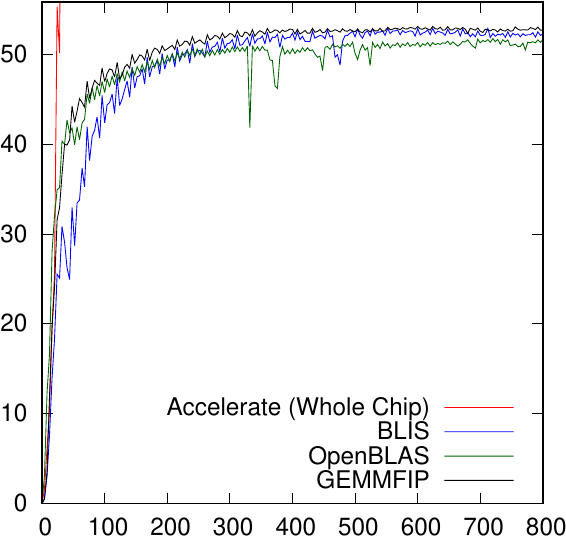}}
\vspace{1.5ex} \\
\rotatebox[origin=c]{90}{GFLOPS/sec w/ $\lda = 2000$} &
\raisebox{-0.49\height}{\includegraphics[width=\graphwidth]{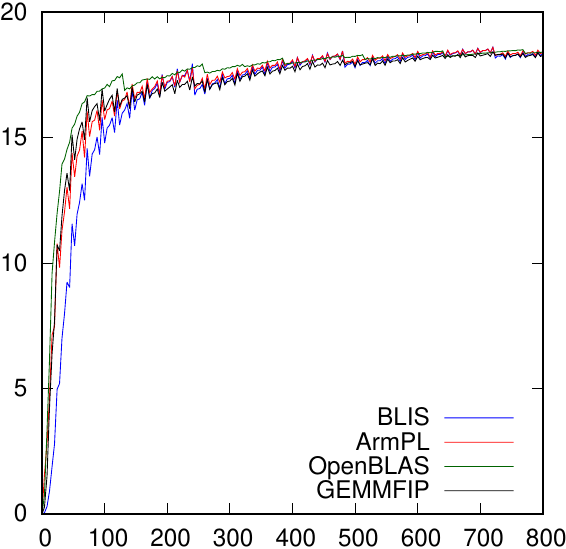}} &
\raisebox{-0.50\height}{\includegraphics[width=\graphwidth]{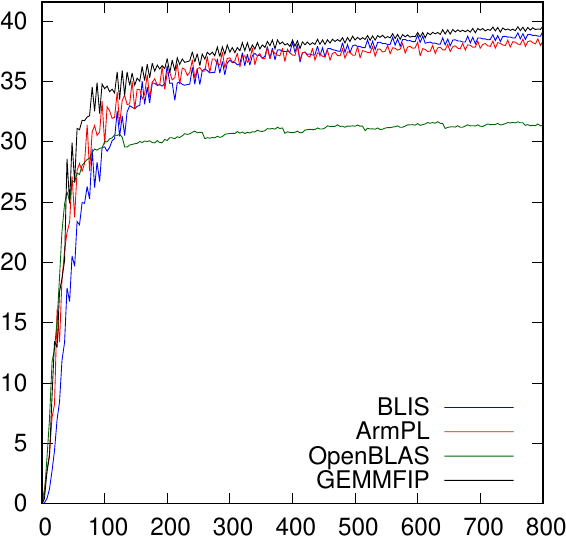}} &
\raisebox{-0.50\height}{\includegraphics[width=\graphwidth]{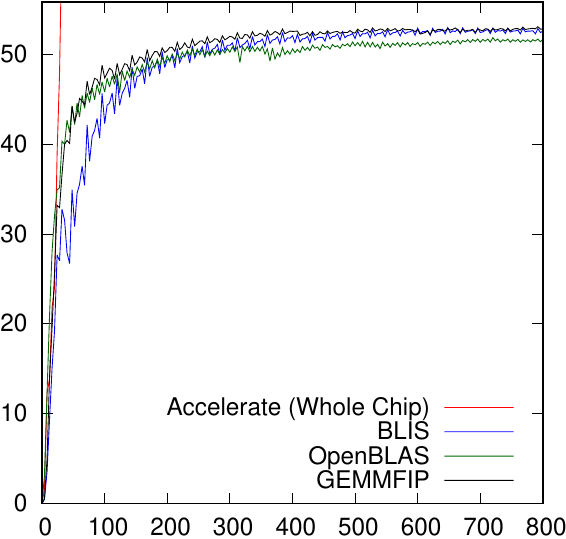}}
\vspace{1.5ex} \\
& $m = n = k$ & $m = n = k$ & $m = n = k$
\end{tabular}
\caption{
Performance on various Arm architectures.  Left: AWS Graviton 2 with \arm.  Middle: AWS Graviton 3 with \armsve\   architecture.   Right: Apple M2 with \arm. }
\label{fig:perf_gemm_ARM}
\end{figure*}